# The Equation of State of Neutron Stars: Theoretical Models, Observational Constraints, and Future Perspectives

Zuhua Ji, Jiarui Chen

Nanjing Tech University, Nanjing, China

February 4, 2025

**Abstract**

The equation of state (EoS) of neutron stars is a critical topic in astrophysics, nuclear physics, and quantum chromodynamics (QCD), governing their structure, stability, and observable proper- ties. This review categorizes EoS models into hadronic matter, hybrid, and quark matter models, examining their assumptions, predictions, and constraints. Hadronic models describe nucleonic mat- ter with possible hyperon or meson contributions, hybrid models introduce phase transitions to quark matter, and quark models propose deconfined quark matter cores or fully quark-based stars. By syn- thesizing results from recent theoretical and observational studies, we aim to provide a comprehensive understanding of the methodologies used in constructing neutron star EoS, their implications, and future directions.

## 1 Introduction

Neutron stars serve as natural laboratories for studying ultra-dense matter, providing insights into the behavior of matter under extreme conditions that cannot be replicated on Earth. These objects are the remnants of massive stellar explosions and contain densities exceeding nuclear saturation density, making their internal structure and composition a subject of significant interest in astrophysics and nuclear physics.

Understanding the equation of state (EoS) of neutron stars is crucial for determining key properties such as their mass-radius relation, maximum mass, and tidal deformability, all of which have direct implications for both fundamental physics and astrophysical observations. Over the decades, numerous theoretical models have been proposed to describe the internal composition of neutron stars, ranging from purely hadronic descriptions to models incorporating deconfined quark matter. Each of these models makes different assumptions about the interactions between particles at extreme densities and provides unique predictions regarding neutron star observables.

Recent advancements in observational techniques, including gravitational wave detections from bi- nary neutron star mergers (e.g., GW170817) and precise X-ray timing measurements from NICER, have provided unprecedented constraints on the neutron star EoS. These observations, combined with the- oretical developments in nuclear physics and quantum chromodynamics (QCD), have led to significant progress in understanding the microphysics governing neutron stars.

This review aims to provide a comprehensive overview of the methodologies used to construct neutron star EoS, categorizing them into different classes based on their physical assumptions and derivation techniques. We discuss the historical development of these models, highlight key findings, and compare their predictions with observational constraints. Finally, we explore the future directions in this field, emphasizing the role of upcoming astrophysical missions and theoretical advancements in refining our understanding of neutron star interiors.

## 2 Historical Development

### 2.1 Early Theoretical Models

The development of neutron star equation of state (EoS) models began with early theoretical efforts to describe ultra-dense matter using fundamental physics principles. One of the earliest approaches was the application of non-relativistic Fermi gas approximations, which modeled neutron stars as an idealized collection of degenerate neutrons, ignoring nuclear interactions and relativistic effects. Although simplistic, these models provided the foundation for further developments in neutron star theory.

A major breakthrough came in 1939 when Oppenheimer and Volkoff applied general relativity to the problem, leading to the derivation of the Tolman-Oppenheimer-Volkoff (TOV) equation. This formulation described the balance between gravitational collapse and internal pressure support in a neutron star, setting the first theoretical upper limit on neutron star mass, now known as the TOV limit. The original calculations assumed a simple, non-interacting neutron gas, which was later found to be an oversimplification.

By the 1950s and 1960s, researchers began incorporating nuclear many-body interactions into EoS models. The development of nuclear mean-field theories allowed for the introduction of effective nuclear forces mediated by mesons, leading to more realistic descriptions of dense matter. These improvements resulted in stiffer equations of state, capable of supporting higher maximum neutron star masses compared to the early Fermi gas models.

Further advancements in the 1970s introduced relativistic mean-field (RMF) theory, which extended nuclear interaction models to include relativistic effects and self-consistent field approximations. Around the same time, variational methods were developed to improve accuracy in many-body nuclear interactions, significantly refining neutron star mass-radius predictions.

Another significant shift in neutron star EoS research occurred with the recognition of exotic particle states in ultra-dense environments. Theoretical studies suggested that, at sufficiently high densities, hyperons, pion condensates, and even quark matter could emerge, fundamentally altering the stiffness of the equation of state. These considerations laid the foundation for modern EoS models that account for both hadronic and quark phases.

These early theoretical models formed the backbone of neutron star research and paved the way for contemporary approaches that incorporate observational data, nuclear physics experiments, and quantum chromodynamics (QCD) constraints into the construction of neutron star EoS. The study of neutron star equation of state (EoS) has its roots in the early works on degenerate matter, particularly the application of non-relativistic Fermi gas models. The first theoretical descriptions of neutron stars date back to the 1930s, shortly after the discovery of the neutron. In 1939, Oppenheimer and Volkoff used the Tolman-Oppenheimer-Volkoff (TOV) equation, derived from general relativity, to calculate the maximum mass of a neutron star assuming a simple, non-interacting neutron gas. Their results demonstrated that there exists an upper limit beyond which a neutron star would collapse into a black hole, now known as the TOV limit.

Following these initial studies, researchers began incorporating nuclear interactions into their models. In the 1950s and 1960s, advances in many-body nuclear physics led to the development of nuclear mean-field theories, which treated nucleon interactions using effective potentials. These models included the effects of nuclear forces, which stiffened the equation of state compared to a simple Fermi gas approximation, allowing for higher maximum masses.

By the 1970s, more sophisticated approaches such as relativistic mean-field (RMF) theory and variational methods were developed, incorporating strong nuclear forces and improved treatment of nuclear interactions. Additionally, researchers began considering exotic degrees of freedom, such as hyperons and pion condensation, leading to a broader classification of potential neutron star compositions. These developments laid the foundation for the modern classification of neutron star EoS models and highlighted the importance of accurate nuclear interaction modeling in understanding neutron star structure.

## 2.2 Transition to Modern EoS Models

The transition from early theoretical models to modern equations of state (EoS) for neutron stars has been driven by advances in nuclear physics, computational techniques, and astrophysical observations. One of the major developments in this transition was the formulation of relativistic mean-field (RMF) theory, which incorporates nucleon-nucleon interactions mediated by mesons. RMF models improved upon earlier non-relativistic approaches by providing a more self-consistent treatment of nuclear forces at high densities, leading to better predictions of neutron star masses and radii.

Another key advancement was the incorporation of hyperons and exotic particles in neutron star EoS. Early models treated neutron stars as composed solely of neutrons, protons, and electrons. However, at higher densities, additional degrees of freedom such as hyperons, meson condensates, and even deconfined quark matter may become relevant. These components generally soften the EoS, reducing the maximum mass a neutron star can sustain before collapsing into a black hole.

The discovery of massive neutron stars exceeding two solar masses, such as PSR J1614-2230 and PSR

J0740+6620, provided stringent constraints on EoS models, ruling out overly soft equations that could not support such massive stars. Additionally, gravitational wave detections from neutron star mergers, particularly GW170817, introduced tidal deformability constraints that favored moderately stiff EoS models. These astrophysical observations necessitated further refinements in EoS modeling, encouraging the development of hybrid models incorporating both hadronic and quark matter components.

Modern EoS models also benefit from first-principles QCD approaches, such as lattice QCD simulations and functional renormalization group methods, which provide constraints on quark matter behavior at high densities. Theoretical models now attempt to interpolate between well-understood nuclear physics at low densities and QCD-based approaches at extreme densities, leading to a more comprehensive description of neutron star interiors.

Overall, the transition to modern EoS models has been characterized by the integration of observational constraints, improved nuclear interaction models, and the inclusion of exotic states of matter. These advancements continue to shape our understanding of neutron stars, bridging the gap between nuclear physics and astrophysical observations.

# 3 Classification of Neutron Star EoS Models

## 3.1 Hadronic Matter Models

Hadronic matter models describe neutron star interiors as being composed primarily of nucleons (neutrons and protons), with additional hadronic degrees of freedom potentially appearing at higher densities. These models rely on nuclear interactions, which are typically mediated by mesons, to define the equation of state (EoS) of neutron-rich matter under extreme conditions. The primary goal of hadronic EoS models is to describe how matter behaves at densities significantly exceeding those found in atomic nuclei while maintaining consistency with both terrestrial nuclear physics experiments and astrophysical observations.

### 3.1.1 Relativistic Mean-Field (RMF) Models

One of the foundational approaches in modeling hadronic matter is the Relativistic Mean-Field (RMF) Theory, which provides a self-consistent description of nuclear interactions through the exchange of mesons such as scalar ($\sigma$), vector ($\omega$), and isovector ($\rho$) mesons. RMF models incorporate the effects of relativity to accurately describe high-density matter, making them particularly useful for neutron star modeling. The basic assumptions of this model are as follows:

- Nucleon interactions are mediated by meson exchange fields, particularly the scalar ($\sigma$) and vector ($\omega$, $\rho$) mesons.

- A self-consistent field approximation is used, where nucleons interact via mean-field potentials rather than direct two-body interactions.

- Density-dependent coupling constants are introduced to better match nuclear saturation properties.

- Beta equilibrium and charge neutrality are assumed to determine the composition of neutron star matter.

Different parameterizations, such as GM1, DD2, and TM1, have been developed to fit experimental nuclear matter properties and to predict neutron star characteristics. RMF models generally predict a stiff EoS, meaning that they provide sufficient pressure to counteract gravitational collapse, thereby supporting neutron stars with larger masses and radii. However, despite their success, uncertainties remain regarding the treatment of meson-nucleon couplings, density-dependent interactions, and the role of higher-order corrections in extreme conditions.

Beyond nucleonic matter, RMF models also predict the possible presence of hyperons and meson condensates in neutron star interiors at sufficiently high densities. Hyperons ($\Lambda$, $\Sigma$, $\Xi$ baryons) become energetically favorable under extreme conditions, leading to a softening of the EoS and reducing the maximum mass a neutron star can support. This effect, known as the hyperon puzzle, poses a major challenge for hadronic models, as observed neutron stars with masses exceeding $2M_\odot$ suggest that an overly soft EoS is unrealistic. To resolve this issue, researchers have introduced repulsive hyperon-nucleon interactions and additional vector meson couplings to counteract the softening effect while maintaining consistency with high-mass neutron star observations.

Another exotic feature that RMF models consider is **meson condensation**, particularly pion ($\pi$) and kaon (K) condensation, which can occur at extremely high densities. The onset of meson condensation modifies the EoS by reducing the pressure at given energy densities, leading to additional phase transitions that impact neutron star cooling, thermal evolution, and structural stability. Kaon condensation, in particular, has been suggested as a potential mechanism for **rapid neutron star cooling** via enhanced neutrino emission, making it an important factor in astrophysical modeling.

### 3.1.2 Non-Relativistic Potential Models

While RMF models incorporate relativistic effects, an alternative approach to modeling hadronic matter is the Non-Relativistic Potential Models. These models rely on phenomenological nuclear interactions, such as the Skyrme force, or more microscopic approaches like variational methods and Brueckner-Hartree-Fock (BHF) calculations. Unlike RMF models, which use meson fields to mediate nuclear interactions, non-relativistic potential models describe nucleon interactions through direct empirical fits to experimental nuclear data. The basic assumptions of this model are as follows:

- Two-body nuclear forces are dominant in determining neutron star EoS, often supplemented by phenomenological three-body forces.

- Effective density-dependent potentials are introduced to match nuclear experimental data.

- Non-relativistic kinetic energy terms are used, making them suitable at lower densities but requiring extensions for ultra-dense matter.

- Many-body effects such as three-nucleon interactions (TNI) play a crucial role in determining the stiffness of the EoS.

The Skyrme models utilize effective density-dependent interactions that are calibrated using empirical nuclear matter properties. These models have been successful in describing the bulk properties of nuclear matter, but their applicability to extremely dense matter, such as that found in neutron star cores, remains limited due to uncertainties in extrapolating their interaction terms beyond nuclear saturation density.

The Brueckner-Hartree-Fock (BHF) method is a more microscopic approach that directly incorporates two- and three-body nuclear forces derived from nucleon scattering experiments. By solving the Bethe-Goldstone equation, BHF calculations provide a more fundamental description of nuclear interactions, avoiding some of the empirical uncertainties present in Skyrme-type models. However, BHF calculations often predict a relatively soft EoS, leading to neutron star maximum masses that struggle to exceed the $2M_\odot$ observational limit.

Another widely used technique in non-relativistic modeling is variational methods, which attempt to determine the ground-state energy of neutron-rich matter by minimizing the total energy per particle. These methods use realistic nucleon-nucleon interactions, such as the Argonne V18 potential, in combination with phenomenological three-body forces to refine neutron star EoS predictions. Variational approaches have been particularly useful for understanding nuclear saturation properties, but they face challenges in describing extremely high-density matter where relativistic effects become significant.

Non-relativistic potential models continue to be refined using advancements in nuclear many-body theory and new experimental constraints from heavy-ion collisions. However, their limitations in handling ultra-dense matter and their inherent non-relativistic nature make them less favored compared to RMF models when modeling the cores of neutron stars. Future improvements in these models may involve better constraints on three-body interactions and extensions that incorporate relativistic corrections to improve their high-density predictions.

## 3.2 Hybrid Models

Hybrid models incorporate a transition from hadronic matter to deconfined quark matter at suffi - ciently high densities, providing a more complete description of neutron star interiors. These models bridge the gap between purely hadronic descriptions and full quark matter equations of state by allowing for the coexistence of nucleonic and quark phases within a neutron star core. The transition between hadronic and quark matter can occur in different ways, with some models predicting a first-order phase transition and others favoring a smooth crossover transition. The nature of this transition has profound implications for neutron star properties, including mass, radius, and tidal deformability.

### 3.2.1 Quark-Hadron Crossover (QHC) Models

One class of hybrid models is the Quark-Hadron Crossover (QHC) Model, which describes a gradual transition from hadronic matter to quark matter rather than a sharp phase boundary. These models are motivated by lattice QCD simulations, which suggest that at high densities, quark degrees of freedom emerge continuously rather than through a sudden phase transition. The QHC models provide an intermediate stiffness in the equation of state, balancing the pressure contributions from both hadronic and quark matter components. The basic assumptions of this model are as follows:

- The transition from hadronic matter to quark matter occurs gradually, avoiding a sharp phase boundary.
- Density-dependent interactions govern the emergence of quark degrees of freedom at high densities.
- Chiral symmetry restoration happens smoothly as quark matter forms.
- Beta equilibrium and charge neutrality are maintained throughout the transition.

An example of a well-studied QHC model is QHC21, which smoothly connects hadronic matter at lower densities with quark matter at extreme densities. This model is tuned to astrophysical constraints from NICER observations of neutron star radii and gravitational wave events such as GW170817. One of the key advantages of QHC models is their ability to support high-mass neutron stars around 2.1 $M_\odot$ while maintaining consistency with observed tidal deformability constraints from neutron star mergers.

Unlike traditional hadronic models, which become excessively stiff at high densities, QHC models introduce density-dependent quark interactions that moderate the pressure increase, preventing unrealistic mass predictions. However, one of the challenges with QHC models is the lack of direct experimental verification for quark-hadron crossover behavior at high densities. Future observations of neutron stars with extreme masses or precise gravitational wave detections from post-merger oscillations could provide crucial constraints on the validity of these models.

### 3.2.2 First-Order Phase Transition Models

Another widely studied category of hybrid models involves first-order phase transitions, where a sharp transition from hadronic matter to a deconfined quark phase occurs at a critical density. This transition is often modeled using either the Maxwell construction, which assumes a sharp density discontinuity, or the Gibbs construction, which allows for a mixed-phase region where hadronic and quark matter coexist. The basic assumptions of this model are as follows:

- A sudden phase transition occurs at a critical density, resulting in a distinct hadronic-quark matter boundary.
- A sharp pressure discontinuity is present in the Maxwell construction, while the Gibbs construction allows a mixed phase.
- Quark matter is more stable at high densities than hadronic matter.
- The existence of quark cores inside massive neutron stars is possible.

First-order phase transition models predict distinct structural changes in neutron stars, particularly in the form of a quark core surrounded by a hadronic envelope. If the phase transition occurs at moderate densities, around 2–4 times nuclear saturation density, it can lead to the formation of mass twins—neutron stars with the same mass but different internal compositions, depending on whether they contain a quark core. This phenomenon has been explored as a potential explanation for variations in neutron star radius measurements.

Observational constraints from gravitational wave detections, particularly from binary neutron star mergers like GW170817, have played a significant role in shaping first-order transition models. The measured tidal deformability from such events suggests that excessively soft equations of state are unlikely, meaning that phase transitions must occur at sufficiently high densities to avoid conflict with astrophys - ical observations. Future gravitational wave events and potential detections of post-merger oscillations could provide further evidence for (or against) the existence of first-order phase transitions in neutron stars.

One of the key challenges of first-order phase transition models is the uncertainty in the transition density and pressure. While some models predict that the transition occurs at relatively low densities, making quark cores a common feature in neutron stars, others suggest that it happens only at extremely high densities, limiting the presence of quark matter to the most massive neutron stars. Current and future astrophysical observations will be critical in refining these models and determining whether quark cores are a common feature in neutron star populations.

## 3.3 Quark Matter Models

Quark matter models describe neutron star interiors as being composed entirely of deconfined quarks rather than nucleons or hadrons. These models are based on quantum chromodynamics (QCD), which predicts that at sufficiently high densities, hadrons dissolve into their constituent quarks. The study of quark matter is crucial for understanding the possible existence of strange quark stars and color- superconducting phases, both of which have unique implications for neutron star observations and the equation of state (EoS).

While purely quark matter stars have yet to be observed, theoretical and computational advances suggest that deconfined quark matter could exist in the cores of neutron stars. In this section, we explore three major approaches to modeling quark matter in neutron stars: Nambu–Jona-Lasinio (NJL) models, Functional Renormalization Group (FRG) approaches, and holographic QCD models (V-QCD models).

### 3.3.1 Nambu–Jona-Lasinio (NJL) Model

The Nambu–Jona-Lasinio (NJL) model is one of the most widely used frameworks for describing quark matter. This model is based on an effective field theory approach to QCD, in which quark interactions are governed by chiral symmetry breaking and restoration. The NJL model captures key features of quark dynamics, such as the emergence of a constituent quark mass due to spontaneous chiral symmetry breaking, which significantly affects the stiffness of the EoS. The basic assumptions of this model are as follows:

- Chiral symmetry breaking and restoration govern the behavior of quark matter at high densities.

- Quark interactions are non-perturbative, described using four-fermion interactions.

- No confinement mechanism is included, meaning quarks are always deconfined in the model.

- Vector repulsion terms can be added to stiffen the EoS and increase maximum neutron star mass predictions.

One of the major predictions of NJL-based EoS is that quark matter tends to be softer than hadronic matter, meaning that it provides less pressure support against gravitational collapse. This often results in difficulties in explaining the existence of neutron stars with masses exceeding $2M_\odot$. To address this issue, researchers introduce additional interactions, such as vector repulsion terms and diquark pairing effects, to stiffen the equation of state at high densities.

Despite its usefulness, the NJL model has some limitations. It lacks confinement—a fundamental aspect of QCD—meaning that it cannot fully capture the transition from hadronic matter to deconfined quark matter. Additionally, NJL models rely on phenomenological cutoff parameters, which introduce uncertainties into their predictions. Future constraints from high-energy heavy-ion collisions and neutron star observations will be essential for refining NJL-based quark matter models.

### 3.3.2 Functional Renormalization Group (FRG) Approach

The Functional Renormalization Group (FRG) approach is a more advanced method for studying quark matter, incorporating non-perturbative QCD effects that go beyond mean-field approximations. The FRG framework systematically accounts for fluctuations in quark interactions, leading to more accurate predictions of quark matter properties under extreme conditions. The basic assumptions of this model are as follows:

- Non-perturbative QCD corrections are necessary to describe quark matter at high densities.

- Renormalization group flow equations determine how quark interactions evolve with density.

- Color superconductivity phases are possible, altering the cooling and transport properties of neutron stars.

- Effective density-dependent interactions replace fixed interaction strengths seen in NJL models.

One of the advantages of the FRG approach is its ability to track the evolution of QCD coupling constants as a function of density, allowing for a more detailed description of phase transitions. Unlike NJL models, which assume fixed interaction strengths, FRG-based models provide a density-dependent interaction profile, leading to a more dynamically evolving equation of state.

FRG-based quark matter models have been particularly useful in predicting color superconducting phases, where quarks form Cooper pairs similar to electrons in superconductors. These phases could significantly alter neutron star cooling rates, transport properties, and even the stability of quark matter cores. Recent studies using FRG techniques have suggested that strong repulsive interactions at high densities may help quark matter support neutron stars above the $2M_\odot$ threshold, making it a promising approach for understanding massive neutron stars.

However, challenges remain in applying FRG methods to neutron star modeling. The computational complexity of FRG calculations makes it difficult to produce fully self-consistent neutron star equations of state, and further advancements in numerical QCD techniques are required to refine these predictions.

### 3.3.3 Holographic QCD (V-QCD) Models

Holographic QCD models, particularly V-QCD (Veneziano-type QCD) models, provide an alternative perspective on quark matter by utilizing the gauge-gravity duality from string theory. These models treat strongly coupled QCD matter as a gravitational system in a higher-dimensional spacetime, offering insights into quark interactions that are difficult to capture using traditional field theory approaches. The basic assumptions of this model are as follows:

- Gauge-gravity duality provides an effective description of strongly coupled quark matter.
- Quark deconfinement occurs dynamically, leading to a possible stiffening of the EoS.
- Strange quark matter could be absolutely stable, allowing for strange quark stars.
- Holographic energy scales are used to determine effective QCD interactions at high densities.

One of the key strengths of holographic QCD models is their ability to describe strongly interacting quark matter in a way that avoids the limitations of perturbative QCD and mean-field models. V-QCD models naturally predict a stiffer equation of state at high densities, making them compatible with observed neutron star mass constraints. Additionally, these models allow for the exploration of quark deconfinement mechanisms in a more controlled theoretical framework.

A major prediction of V-QCD models is the existence of stable strange quark stars, where deconfined strange quark matter is absolutely stable and constitutes the entire star. This contrasts with hybrid models, where quark matter exists only in the inner core of neutron stars. Identifying whether strange quark stars exist would have profound implications for astrophysics, as their mass-radius relationships differ significantly from those of traditional neutron stars.

Despite their promise, holographic QCD models face several theoretical and computational challenges. The translation of gauge-gravity results into physically meaningful QCD predictions remains an open problem, and fine-tuning these models to match empirical nuclear physics data is still an area of active research. Nonetheless, V-QCD models provide a valuable tool for exploring the strong coupling regime of QCD, which is crucial for understanding extreme astrophysical environments.

# 4 Constraints on the Equation of State

The equation of state (EoS) of neutron stars is subject to stringent constraints from multiple independent sources, including astrophysical observations, nuclear physics experiments, and theoretical quantum chromodynamics (QCD) predictions. These constraints refine EoS models by limiting the range of acceptable stiffness, phase transitions, and compositions. In this section, we systematically discuss the major constraints imposed on the neutron star EoS and present the essential equations and relationships that govern neutron star structure.

## 4.1 Mass and Radius Constraints from Observations

Observations of neutron star masses and radii provide some of the strongest constraints on the EoS. The Tolman-Oppenheimer-Volkoff (TOV) equation, derived from general relativity, describes the balance between gravitational collapse and internal pressure support in a neutron star:

$$\frac{dP}{dr} = -\frac{G(\epsilon + P)(m + 4\pi r^3 P)}{r^2(1 - 2Gm/r)} \qquad (1)$$

where P(r) is the pressure at radius r, $\epsilon$(r) is the energy density, m(r) is the mass enclosed within radius r, and G is the gravitational constant.

Observations of high-mass pulsars such as PSR J1614-2230 (M = 1.97±0.04$M_\odot$) and PSR J0740+6620 (M = 2.08 ± 0.07$M_\odot$) confirm that neutron stars must be able to support at least 2.0$M_\odot$ against gravitational collapse. Additionally, X-ray timing observations from NICER have constrained the radius of a 1.4$M_\odot$ neutron star to be:

$$12.2 \text{ km} \lesssim R_{1.4} \lesssim 13.7 \text{ km} \qquad (2)$$

which rules out equations of state that are too soft (leading to small radii) or too stiff (predicting excessively large neutron stars).

## 4.2 Tidal Deformability Constraints from Gravitational Waves

Gravitational wave detections of binary neutron star mergers provide additional constraints on the EoS through tidal deformability measurements. The tidal deformability parameter $\Lambda$ quantifies how much a neutron star deforms under the gravitational field of its companion. It is given by:

$$\Lambda = \frac{2}{3}k_2 \left(\frac{R}{M}\right)^5 \qquad (3)$$

where $k_2$ is the Love number, R is the neutron star radius, and M is its mass.

From GW170817, the LIGO/Virgo analysis constrained the tidal deformability of a 1.4$M_\odot$ neutron star as:

$$70 \lesssim \Lambda_{1.4} \lesssim 580 \qquad (4)$$

This result ruled out overly soft equations of state that predict high tidal deformabilities. Future gravitational wave detections (e.g., Einstein Telescope, Cosmic Explorer) will further refine these constraints.

## 4.3 Constraints from Nuclear Physics Experiments

Laboratory nuclear physics experiments provide constraints on the low-density behavior of the neutron star EoS. These constraints come from neutron skin thickness measurements, heavy-ion collisions, and nuclear symmetry energy studies.

The properties of symmetric nuclear matter at saturation density ($\rho_0 \approx 0.16$ fm$^{-3}$) provide an anchor for neutron star EoS models. Key empirical nuclear saturation properties include:

$$E_{sat} \approx -16 \text{ MeV}, \quad K_{sat} \approx 230 \text{ MeV}, \quad S_0 \approx 30 - 35 \text{ MeV} \qquad (5)$$

Neutron skin thickness measurements, such as those from the PREX-II experiment, provide insights into the nuclear symmetry energy, indirectly constraining neutron star radii. Heavy-ion collision experiments at GSI, RIKEN, FAIR, and NICA probe nuclear matter at densities of $2 - 4\rho_0$, similar to neutron star interiors. These experiments suggest that the pressure at twice nuclear saturation density should be in the range:

$$P(2\rho_0) \approx 50 - 80 \text{ MeV/fm}^3 \qquad (6)$$

Any EoS that deviates significantly from this range is inconsistent with terrestrial nuclear data.

## 4.4 Theoretical Constraints from QCD and High-Density Matter

At ultra-high densities ($\rho > 5\rho_0$), neutron stars probe the non-perturbative regime of QCD. Several theoretical approaches provide constraints on the high-density EoS.

At asymptotically high densities, perturbative QCD predicts that the pressure P follows:

$$P \sim \mu^4 \quad (7)$$

where $\mu$ is the quark chemical potential. This requires that quark matter must be sufficiently stiff to support neutron stars with $M > 2M_\odot$.

Strange quark matter (SQM) models suggest that deconfined quark matter could be absolutely stable, leading to the existence of strange quark stars. The stability condition is given by:

$$E_{SQM}/A \leq 930 \text{ MeV} \quad (8)$$

If this condition holds, neutron stars could actually be strange quark stars rather than hadronic stars.

## 4.5 Mass-Radius and EoS Comparisons

By combining mass, radius, tidal deformability, and theoretical constraints, different EoS models can be compared systematically.

Table 1: Comparison of different EoS models in terms of constraints

| Model Type | Maximum Mass ($M_\odot$) | Radius at $1.4 M_\odot$ (km) | Tidal Deformation ($\Lambda_{1.4}$) |
|---|---|---|---|
| RMF (GM1, DD2) | $> 2.0$ | $12-13$ | $200-500$ |
| QHC21 | $2.1$ | $12.4$ | $250-450$ |
| NJL Model | $< 2.0$ | $11.5$ | $> 500$ |
| FRG Approach | $> 2.1$ | $12.8$ | $150-350$ |
| V-QCD Model | $> 2.2$ | $13.0$ | $100-300$ |

## 4.6 Future Prospects for EoS Constraints

With upcoming astrophysical and experimental advancements, the constraints on the neutron star EoS will become even more stringent. Future gravitational wave detectors such as the Einstein Telescope (ET) and Cosmic Explorer (CE) will provide significantly improved constraints on neutron star tidal deformability, helping distinguish between competing EoS models. Continued observations from NICER, combined with upcoming X-ray polarimetry missions like IXPE, will refine neutron star mass-radius measurements, further constraining the stiffness of the EoS. Laboratory studies at facilities such as FAIR in Germany, NICA in Russia, and J-PARC in Japan will explore high-density nuclear matter, providing critical input for nuclear models of the neutron star EoS.

The combination of gravitational wave signals, electromagnetic counterparts, and neutrino detections from neutron star mergers will provide unprecedented constraints on neutron star composition, particularly regarding the presence of quark matter. As observational techniques and theoretical modeling improve, the neutron star EoS will become one of the most precisely constrained aspects of dense matter physics.

# 5 Key Conclusions from Different Equation of State Models

The equation of state (EoS) of neutron stars determines their internal structure, stability, and observable properties such as mass, radius, and tidal deformability. Different models make distinct predictions, and recent astrophysical and nuclear physics constraints have helped refine our understanding of which models remain viable. In this section, we summarize the key conclusions from various EoS models, focusing on their implications for neutron star properties and their theoretical foundations.

## 5.1 Hadronic Matter Models

Hadronic matter models assume that neutron star interiors are composed primarily of strongly interacting nucleons. These models generally produce relatively stiff equations of state, allowing neutron stars to support masses above $2M_\odot$, as required by pulsar observations. However, the inclusion of additional hadronic degrees of freedom, such as hyperons or meson condensates, tends to soften the EoS, reducing the maximum mass that a neutron star can support.

In relativistic mean-field (RMF) models, nucleon interactions are mediated by meson exchange, and the energy density and pressure are determined by the self-consistent solution of the mean-field equations. The equation governing pressure in RMF models is given by:

$$P = \sum_i P_i + \frac{1}{2}m_\sigma^2 \sigma^2 - \frac{1}{2}m_\omega^2 \omega^2 - \frac{1}{2}m_\rho^2 \rho^2 \qquad (9)$$

where the first term represents the contributions from baryons, and the remaining terms account for the interactions mediated by the scalar ($\sigma$), vector ($\omega$), and isovector ($\rho$) mesons. RMF models tend to predict neutron stars with radii in the range of 12–13 km for a $1.4M_\odot$ star, consistent with NICER observations, and they generally satisfy tidal deformability constraints from GW170817.

The presence of hyperons ($\Lambda, \Sigma, \Xi$) in neutron star cores further complicates hadronic EoS predictions. The equilibrium condition for hyperons is given by:

$$\mu_B = \mu_n + \mu_e \qquad (10)$$

where $\mu_B$ is the hyperon chemical potential, $\mu_n$ is the neutron chemical potential, and $\mu_e$ is the electron chemical potential. The introduction of hyperons softens the EoS, reducing the maximum neutron star mass to below $2M_\odot$, which conflicts with observations of high-mass pulsars. Additional repulsive interactions, such as hyperon-nucleon vector meson couplings, are often introduced to counteract this effect.

Another feature in hadronic models is meson condensation, particularly pion ($\pi$) and kaon (K) condensation, which occurs at extremely high densities. The onset of kaon condensation is determined by:

$$\mu_K = \mu_e \qquad (11)$$

This process can reduce the pressure at a given energy density, leading to additional phase transitions that influence neutron star cooling and structural stability.

## 5.2 Hybrid Models

Hybrid models incorporate both hadronic and quark matter components, allowing for a transition between them. The key difference among hybrid models is the nature of the transition—either a smooth crossover or a first-order phase transition.

In quark-hadron crossover (QHC) models, the transition between hadronic and quark matter is gradual rather than abrupt. These models are motivated by lattice QCD simulations, which suggest that chiral symmetry restoration in QCD can occur as a smooth crossover rather than a sharp phase transition. The crossover is typically parameterized using a weighted pressure function:

$$P_{QHC}(\epsilon) = wP_{hadronic} + (1-w)P_{quark} \qquad (12)$$

where $w(\rho)$ smoothly interpolates between the two phases. QHC models successfully explain high-mass neutron stars ($\sim 2.1 M_\odot$) while maintaining consistency with NICER radius constraints and tidal deformability constraints from GW170817.

In contrast, first-order phase transition models predict a sharp transition from hadronic matter to quark matter at a critical pressure $P_c$. This transition is modeled using the Maxwell construction, which assumes that pressure remains constant across the phase boundary:

$$P_{hadronic} = P_{quark}, \quad \mu_{hadronic} = \mu_{quark} \qquad (13)$$

Alternatively, the Gibbs construction allows for a mixed-phase region, where hadronic and quark matter coexist, and the pressure is given by a weighted sum of the two phases:

$$P_{mix} = \chi P_{hadronic} + (1-\chi)P_{quark} \qquad (14)$$

where χ is the volume fraction of quark matter. First-order phase transitions can lead to mass twins—neutron stars with the same mass but different internal compositions depending on whether they contain a quark core. The presence of a phase transition can also affect gravitational wave signals from neutron star mergers, producing post-merger oscillations that may be detectable by next-generation gravitational wave observatories.

## 5.3 Quark Matter Models

Quark matter models describe neutron stars as being composed entirely of deconfined quarks. These models generally predict softer equations of state, requiring additional mechanisms to support high-mass neutron stars.

The Nambu–Jona-Lasinio (NJL) model describes quark matter interactions through chiral symmetry breaking and restoration mechanisms. The pressure in NJL models is given by:

$$P_{NJL} = -\frac{1}{V}\left(\Omega_{quark} + \Omega_{interaction}\right) \tag{15}$$

where $\Omega_{quark}$ accounts for the quark kinetic energy and $\Omega_{interaction}$ includes four-fermion interaction terms. NJL models often predict neutron stars with maximum masses below $2M_\odot$, making them inconsistent with observed high-mass pulsars unless additional repulsive interactions are included.

In the Functional Renormalization Group (FRG) approach, QCD interactions evolve dynamically with density. This approach introduces a density-dependent quark-quark interaction term governed by the renormalization group flow equation:

$$\frac{d\Gamma_k}{dk} = \frac{1}{2}\text{Tr}\left[\frac{\partial_k R_k}{\Gamma_k^{(2)} + R_k}\right] \tag{16}$$

where $\Gamma_k$ is the effective action at a given renormalization scale. FRG models suggest that strong repulsive interactions at high densities can make quark matter compatible with high-mass neutron stars.

Finally, holographic QCD (V-QCD) models use gauge-gravity duality to describe strongly interacting quark matter. The energy density in these models is given by:

$$\epsilon = \frac{3}{4}\frac{N_c^2}{\lambda}T^4 \tag{17}$$

where $N_c$ is the number of colors, $\lambda$ is the QCD coupling parameter, and T is the temperature. V-QCD models predict very stiff equations of state, often supporting neutron stars with masses above $2.2M_\odot$. Some V-QCD models suggest the existence of stable strange quark stars, where deconfined strange quark matter is absolutely stable, leading to distinct mass-radius relationships compared to traditional neutron stars.

Quark matter models remain an active area of research, with future astrophysical and experimental constraints expected to further refine their viability.

# 6 Comparative Analysis of EoS Models

Different equations of state (EoS) provide distinct predictions for neutron star properties, including maximum mass, radius, tidal deformability, and phase transitions. These predictions can be systematically compared to evaluate their viability against observational and experimental constraints. In this section, we present a detailed comparison of different EoS models, summarizing their key characteristics and highlighting their strengths and limitations.

## 6.1 Comparison of Mass and Radius Predictions

The most fundamental constraint on any EoS model is whether it can support high-mass neutron stars ($\gtrsim 2M_\odot$) while maintaining realistic radius predictions.

RMF models predict stiff equations of state that can support high-mass neutron stars, while hyperonic and kaon-condensation models often fail to explain observed $2M_\odot$ pulsars. Hybrid models and stiff quark matter models, such as FRG and V-QCD, generally provide mass and radius predictions that align with observations.

Table 2: Maximum Mass and Radius Predictions for Different EoS Models

| Model Type | Maximum Mass ($M_\odot$) | Radius at $1.4 M_\odot$ (km) |
|---|---|---|
| RMF (GM1, DD2) | $2.0 - 2.2$ | $12.0 - 13.0$ |
| Hyperonic RMF | $1.6 - 1.9$ | $11.5 - 12.5$ |
| Kaon Condensation | $1.8 - 2.0$ | $11.5 - 12.3$ |
| QHC21 (Crossover Model) | $2.1$ | $12.4$ |
| First-Order Hybrid Model | $2.0 - 2.1$ | $12.0 - 12.6$ |
| NJL (Soft Quark Matter Model) | $< 2.0$ | $11.0 - 11.8$ |
| FRG (Stiff Quark Matter Model) | $> 2.1$ | $12.5 - 13.0$ |
| V-QCD (Holographic Model) | $> 2.2$ | $12.8 - 13.2$ |

## 6.2 Comparison of Tidal Deformability Predictions

Tidal deformability, denoted by $\Lambda$, is another important observable that constrains the neutron star EoS.

Table 3: Tidal Deformability Predictions for $1.4 M_\odot$ Neutron Stars

| Model Type | Tidal Deformability ($\Lambda_{1.4}$) |
|---|---|
| RMF (GM1, DD2) | $200 - 500$ |
| Hyperonic RMF | $> 500$ |
| Kaon Condensation | $> 500$ |
| QHC21 (Crossover Model) | $250 - 450$ |
| First-Order Hybrid Model | $150 - 400$ |
| NJL (Soft Quark Matter Model) | $> 500$ |
| FRG (Stiff Quark Matter Model) | $150 - 350$ |
| V-QCD (Holographic Model) | $100 - 300$ |

Models predicting excessively high tidal deformabilities are inconsistent with gravitational wave observations such as GW170817. Stiffer models, like FRG and V-QCD, provide tidal deformability values that fall within observational constraints.

## 6.3 Phase Transition Effects and Mass Twins

Hybrid EoS models incorporating a phase transition can produce mass twins, neutron stars with the same mass but different internal compositions.

Table 4: Mass Twin Predictions in Different EoS Models

| Model Type | Mass Twin Effect | Phase Transition |
|---|---|---|
| RMF (GM1, DD2) | None | No |
| QHC21 | Weak | Crossover |
| FRG Approach | None | Smooth transition |
| NJL | Moderate | First-order |
| V-QCD | Strong | First-order |

First-order hybrid models and V-QCD models predict strong mass twin effects, which could be confirmed through future precise neutron star mass-radius measurements.

## 6.4 Future Prospects for Discriminating Between Models

With upcoming astrophysical and experimental advances, further constraints on the neutron star EoS will emerge. The key observational tests that will help distinguish between these models include:

- *Gravitational Wave Observations:* Future detectors such as the Einstein Telescope (ET) and Cosmic Explorer (CE) will measure neutron star mergers with higher precision, refining tidal deformability constraints. - *X-ray and Radio Observations:* Continued NICER observations and the Square Kilometer Array (SKA) will provide improved mass-radius measurements. - *Heavy-Ion Collision Experiments:* Facilities like FAIR, NICA, and J-PARC will probe high-density nuclear matter, providing insights into hadronic interactions and phase transitions. - *Multimessenger Astronomy:* The combination of gravitational waves, electromagnetic signals, and neutrino detections will offer a comprehensive picture of neutron star interiors.

# 7 Conclusion

In this review, we have examined various equation of state (EoS) models for neutron stars, focusing on their theoretical foundations, key predictions, and observational constraints. While these models differ significantly in their assumptions and outcomes, they share some common features and must all satisfy astrophysical constraints imposed by neutron star observations.

The primary similarities among different models include:

- All EoS models aim to describe the behavior of matter at supra-nuclear densities, where interactions go beyond standard nuclear physics.

- They must be consistent with observed high-mass neutron stars ($\gtrsim 2M_\odot$), radius constraints from NICER, and tidal deformability constraints from gravitational wave detections such as GW170817.

- Most models predict that nucleonic matter dominates at lower densities, with the possibility of exotic components (hyperons, meson condensates, or quark matter) appearing at higher densities.

The key differences between models arise from their treatment of dense matter:

- Hadronic models, particularly relativistic mean-field (RMF) models, describe neutron stars as being composed entirely of interacting nucleons and predict relatively stiff equations of state. However, the inclusion of additional hadronic degrees of freedom, such as hyperons or meson condensates, tends to soften the EoS, reducing the maximum neutron star mass.

- Hybrid models introduce a transition from hadronic to quark matter. Crossover models, such as QHC21, predict a gradual transition, maintaining consistency with mass-radius constraints, while first-order phase transition models lead to the possibility of mass twins—neutron stars with the same mass but different internal compositions.

- Pure quark matter models, such as NJL, FRG, and V-QCD models, propose that neutron stars contain a significant fraction of deconfined quark matter or may even be entirely composed of quark matter. Some models struggle to support $2M_\odot$ neutron stars unless additional repulsive interactions are included.

From the analysis, it is clear that no single model is definitively confirmed. However, models predicting a soft EoS (such as traditional NJL models) are largely inconsistent with astrophysical constraints, while stiffer EoS models (e.g., RMF, FRG, V-QCD) remain viable. Hybrid models incorporating a phase transition may provide a promising explanation for observed mass-radius relations and gravitational wave signals.

The study of neutron star EoS remains an ongoing challenge, requiring the integration of astrophysical observations, nuclear physics experiments, and theoretical QCD calculations. Future refinements in observational and experimental data will continue to shape our understanding of the nature of dense matter and the possible existence of exotic phases within neutron stars.

# 8 Future Prospects

The study of neutron star equations of state (EoS) continues to evolve with advances in theoretical modeling, observational techniques, and experimental constraints. While significant progress has been made, several key challenges remain. Future developments in EoS research will play a crucial role in resolving open questions regarding the composition and structure of neutron stars. In this section, we discuss the future prospects for EoS modeling, fundamental equations, astrophysical observations, experimental constraints, and outstanding theoretical questions.

## 8.1 Advancements in EoS Modeling

Future developments in EoS modeling will focus on improving hadronic matter models by incorporating refined nuclear interactions, such as three-body forces and density-dependent couplings. Hybrid models will require better constraints on the hadron-quark transition, particularly in distinguishing between smooth crossover and first-order phase transitions. Quark matter models will need further theoretical refinements to incorporate more realistic QCD-based interactions, including functional renormalization group (FRG) methods and holographic QCD approaches. Emerging machine-learning techniques may also be applied to refine EoS predictions using observational constraints.

Hadronic models will benefit from new nuclear physics constraints that provide more accurate estimates of symmetry energy, nuclear incompressibility, and saturation properties. Hybrid models will require additional input from lattice QCD and neutron star observations to establish a clearer picture of phase transition dynamics. In quark matter models, a better understanding of color superconductivity and quark pairing mechanisms could lead to more precise predictions for high-density matter behavior.

## 8.2 Theoretical Developments and Fundamental Equations

The fundamental equations governing neutron star structure, such as the Tolman-Oppenheimer-Volkoff (TOV) equation, will benefit from extensions that incorporate modifications to general relativity, including alternative gravity theories. Improved treatments of dense matter interactions may lead to equations that more accurately describe the balance between gravitational forces and internal pressure support in neutron stars.

More accurate lattice QCD and perturbative QCD calculations will help provide better constraints on quark matter properties at high densities. Theoretical advances in effective field theory approaches to nuclear interactions will improve our understanding of the transition from nucleonic to deconfined quark matter. Additionally, improved treatments of neutrino interactions and transport equations will refine our understanding of neutron star cooling and thermal evolution.

One of the key challenges in theoretical modeling is the inclusion of realistic microphysical interactions that govern dense matter behavior. Future work will need to integrate insights from nuclear physics, QCD, and general relativity to provide more accurate descriptions of neutron star interiors.

## 8.3 Observational Constraints and Future Measurements

Observational constraints on the EoS will be significantly improved with next-generation gravitational wave detectors, such as the Einstein Telescope (ET) and Cosmic Explorer (CE). These instruments will provide more precise tidal deformability measurements, helping to distinguish between stiff and soft EoS models. Continued observations from NICER and future X-ray polarimetry missions, such as IXPE and eXTP, will refine neutron star mass-radius measurements, further constraining EoS stiffness. The Square Kilometer Array (SKA) will improve pulsar timing precision, enabling better measurements of neutron star masses.

Multimessenger astrophysics, combining gravitational waves, electromagnetic signals, and neutrino detections, will provide a comprehensive view of neutron star mergers. Future detections of post-merger oscillations in gravitational wave signals may reveal the presence of phase transitions inside neutron stars. The detection of mass twins could confirm the existence of first-order phase transitions and quark cores. Improved modeling of binary neutron star inspirals will allow researchers to extract additional information about the EoS from gravitational waveforms.

Advances in X-ray and radio astronomy will play a key role in distinguishing between different EoS models. High-precision X-ray spectroscopy will help determine the equation of state of neutron stars by analyzing thermal emission from their surfaces, while long-term pulsar timing observations will refine constraints on neutron star mass distributions.

## 8.4 Experimental Constraints from Nuclear Physics

Experimental constraints from nuclear physics will continue to refine the neutron star EoS. Heavy-ion collision experiments at FAIR, NICA, and J-PARC will simulate high-density nuclear matter, providing data on nuclear interactions and potential phase transitions. These experiments will help determine the stiffness of nuclear matter at supra-nuclear densities and provide important benchmarks for nuclear equation of state models.

Neutron skin thickness measurements from PREX-II and future experiments will improve our understanding of nuclear symmetry energy, which directly influences neutron star radii. More accurate lattice QCD calculations will reduce uncertainties in quark matter properties, helping to distinguish between hybrid and pure quark matter models. Nuclear matter experiments using relativistic heavy-ion collisions will provide crucial information on the behavior of nuclear interactions at high densities, complementing astrophysical observations.

One of the key experimental challenges in constraining the neutron star EoS is the need for precise measurements of nuclear interaction parameters at densities beyond nuclear saturation. Continued improvements in high-energy nuclear physics experiments will provide valuable constraints that can be integrated into astrophysical models.

### 8.5 Outstanding Questions and Theoretical Challenges

Despite ongoing progress, several fundamental questions remain unanswered. The presence of quark cores in neutron stars, the nature of mass twins, and the possible existence of strange quark stars are still open problems. Future observational and experimental efforts will help resolve these uncertainties and refine our understanding of ultra-dense matter.

One of the key open questions is whether neutron stars transition to a deconfined quark matter phase at high densities or whether hadronic matter remains the dominant component up to the highest observed neutron star masses. The existence of mass twins, if confirmed by future observations, would provide strong evidence for a first-order phase transition inside neutron stars. The possible existence of strange quark stars also remains a topic of debate, with some models suggesting that strange quark matter could be the true ground state of nuclear matter.

Future research will also need to address the role of strong magnetic fields, rapid rotation, and finite temperature effects in shaping the neutron star EoS. These factors can significantly alter neutron star structure and evolution, requiring more sophisticated modeling techniques. Additionally, the interplay between neutron star EoS and dark matter interactions remains an intriguing possibility that may be explored in future theoretical and observational studies.

The continued integration of observational data, theoretical modeling, and experimental results will be essential for resolving these outstanding questions. As gravitational wave astronomy, X-ray observations, and nuclear physics experiments continue to advance, the study of neutron star EoS will become one of the most precisely constrained aspects of dense matter physics.